\documentclass[preprint]{revtex4-1}
\usepackage{natbib,hyperref}
\usepackage{epsfig}
\usepackage[usenames,dvipsnames,svgnames,table]{xcolor}
\usepackage{amssymb}
\usepackage{amsmath}

\newcommand{\eodim}{B(\text{D})^{}}

\newcommand{\eotri}{B(\text{T})^{}}
\newcommand{\eohe}{B(^3\text{He})^{}}

\newcommand{\rgm}{$\mathbb{R}$GM}

\newcommand{\mn}{m_N}
\newcommand{\mpi}{$m_\pi$}

\newcommand{\ie}{\textit{i.e.}}
\newcommand{\eg}{\textit{e.g.}\;}
\newcommand{\kty}{k_\text{\tiny typ}}
\newcommand{\app}{a^C_{pp}}

\newcommand{\eftnopi}{EFT(\mbox{$\pi\text{\hspace{-6.3pt}/}$)}}

\begin{document}
\title{The Coulomb interaction in Helium-3:\\ Interplay of strong short-range and weak long-range potentials} 
\author{J.~Kirscher}
\email{j.kirscher@mail.huji.ac.il}
\author{D.~Gazit}
\email{doron.gazit@phys.huji.ac.il}
\address{Racah Institute of Physics, The Hebrew University, Jerusalem 91904, Israel}
\begin{abstract}
Quantum chromodynamics and the electroweak theory at low energies are prominent instances of the combination of a short-range and a long-range interaction.
For the description of light nuclei, the large nucleon-nucleon scattering lengths produced by the strong interaction, and the reduction of the weak interaction to the Coulomb potential, play a crucial role.
Helium-3 is the first bound nucleus comprised of more than one proton in which this combination of forces can be studied.

We demonstrate a proper renormalization of Helium-3 using the pionless effective field theory as
the formal representation of the nuclear regime as strongly interacting fermions.
The theory is found consistent at leading and next-to-leading order without isospin-symmetry-breaking 3-nucleon interactions and a \textit{non-perturbative} treatment of the Coulomb interaction.
The conclusion highlights the significance of the regularization method since a comparison to previous work is contradictory if the difference in those methods is not considered.

With a \textit{perturbative} Coulomb interaction, as suggested by dimensional analysis, we find the Helium-3 system properly renormalized, too.

For both treatments, renormalization-scheme independence of the effective field theory is demonstrated by regulating the potential and a variation of the associated cutoff.
\end{abstract}
\maketitle
\section{Introduction}
Quantum chromodynamics (QCD) is non-perturbative at low energies where it is
characterized by a scale separation. These two facts facilitate an approximate solution of low-energy
QCD, \ie, nuclear physics, with renormalization group (RG)
and effective field theory (EFT) techniques~\cite{1979PhyA...96..327W,1990PhLB..251..288W,1991NuPhB.363....3W,1984AnPhy.158..142G}.

The strongly interacting character of QCD is of particular interest at very low-energies. There,
the nuclear regime can be described solely by nucleons interacting at the same space-time point, since the excitation 
of other degrees of freedom is dynamically forbidden. 
This ``pionless'' EFT (\eftnopi) of nuclear physics is characterized by the large nucleon-nucleon ($NN$) 
scattering lengths relative to the effective range of the nuclear force, as 
indicated by the unnaturally small deuteron binding energy~\cite{Bethe146,Bethe176}. \eftnopi~reproduces Bethe's 
effective range theory~\cite{PhysRev.76.38} as an expansion of the $NN$ amplitude about a non-trivial fixed point
of the RG, \ie, unitary fixed point. Thereby, it describes strongly interacting fermionic systems with infinite
scattering lengths (original formulation~\cite{vanKolck1999273,Kaplan:1998tg} and RG emphasis~\cite{Birse1999169}) at its leading order (LO).
For its usefulness in larger systems, \eftnopi~includes a 3-body contact interaction at LO,
abandoning the naturalness assumption and na\"ive dimensional analysis for this operator.
The enhancement of the \mbox{3-body-contact} interaction
is related to a limit cycle found in the RG analysis of the
triton~\cite{PhysRevLett.82.463} and thereby also an expression
of the specific regularization that facilitated the limit cycle and gave it its shape.
The appearance of 3-body bound states at threshold associated with the limit cycle is a reminiscence of Efimov physics in
the unitary limit~\cite{1970PhLB...33..563E}.
Na\"ive power counting fails due to proximity to the non-trivial unitary fixed point.
Perturbations at higher orders include effective-range corrections, as well as relativistic effects.
Additional counter terms, needed to renormalize 3-nucleon forces in the triton,
appear only at next-to-next-to leading order (NNLO)~\cite{Bedaque:1999ve,Griesshammer:2005ga}.

The accidental separation of scales inducing the above \eftnopi~power counting is realized, in particular, between the large scattering lengths and the mass of the pion, as the lightest mesonic degree of freedom.
The scale $m_\pi$ suggests that the \eftnopi~approach is limited to light nuclei ($A\lesssim 4$),
since the scale set by the binding momentum in heavier nuclei is of the order of the pion mass $m_\pi$,
implying their mandatory explicit inclusion in the theory. As QCD generates this large scattering length, its effective potential
between nuclei is, relative to the scattering length, short-ranged.

The existence of a long-range repulsive interaction, namely the Coulomb exchange between protons, complicates the description of nuclei relative to
generic systems in the unitary limit which are canonically analyzed solely by means of short-range potentials.

The electromagnetic interaction 
plays an important role in many applications of nuclear physics. 
The importance of the Coulomb force, in particular,
rises as the momentum scale decreases. This expectation follows from the na\"ive scale of the effect, $\frac{\alpha \mn}{Q}$,
which becomes significant for momenta $Q\lesssim 10~$MeV (nucleon mass $\mn$, and fine structure constant $\alpha$).
Low-energy proton-proton scattering, \eg,~\cite{Kong:1999tw}, is just one example representing a wealth of
bound and scattering few-body systems where this enhancement requires
an accurate description of the Coulomb force.

The conditions for a perturbative treatment for $A>2$ are non-trivial to assess. The complexity is apparent
in the seemingly contradictory observations of Coulomb effects only distorting the asymptotic wave
function in the analysis of low-energy proton-proton scattering~\cite{PhysRev.75.1637,RevModPhys.22.77},
on the one hand, while on the other hand a new counter-term\footnote{This term is small relative to its analog in the neutron-proton channels which reconciles the two statements, eventually.} is needed to fit the low-energy proton-proton scattering
length~\cite{Kong2000137}.
The Helium-3 (helion) nucleus constitutes an ideal (and maybe the only, if
\eftnopi~is inapplicable in heavier nuclei) nuclear specimen to study
the effect of a long-range force on the dynamics of otherwise short-range interacting fermions.

The problem is of universal character as we understand nuclear physics as one representative of a class of theories
which sit in the vicinity of a ``critical manifold'' in the space of 2 and 3-body coupling constants: an infrared
fixed point and a limit cycle.
In the presence of long-range forces, RG properties are changed
significantly~\cite{PhysRevLett.29.917,PhysRevC.67.064006}. Thus, nuclear physics
at very low energies, characterized by the strong short-range forces of the unitary fixed point,
and weak Coulomb long-range forces, provides a model, as well as a physical system.

\eftnopi~studies of Helium-3 with cutoff variation have been first accomplished in
2010~\cite{2010JPhG...37j5108A,PhysRevC.83.064001,FewBodySyst.54.1},
validating the LO power counting. 
The recent helion analysis~\cite{PhysRevC.89.064003} is remarkable for its unnatural promotion of
another 3-nucleon operator.
From a cutoff dependence of the helion binding energy ($\eohe$) at
next-to-leading order (NLO), the enhancement of an iso-asymmetric 3-nucleon interaction to NLO was inferred.
The nuclear bound state problem was formulated and solved, in that work, using Lippmann-Schwinger integral
equations and the Coulomb interaction was included non-perturbatively.
Here we use the analogous Schr\"odinger approach to solve the nuclear problem,
which effectively imposes a cutoff on the potential and in that differs from the method employed in~\cite{PhysRevC.89.064003}.
In our scheme, results do not exhibit a strong cutoff dependence.

In this article, we present an \eftnopi~calculation at NLO, including the renormalized Coulomb interaction.
We outline a consistent scheme to study \eftnopi~at NLO, and show that it is equivalent to the
distorted-wave Born approximation. 
The predictive power is demonstrated in the Helium-3 system.  
Moreover, we show that the results are consistent with a perturbative treatment of the Coulomb force in this nucleus.
\section{Perturbative Next-To-Leading order \eftnopi}
In this section, we summarize the power counting of the \eftnopi~up to NLO. Power-counting schemes, in general,
order contributions of the arbitrarily complicated interaction operators built of the nucleon field $N$ and
its derivatives, which appear in the effective Lagrangian, to the
scattering amplitude. We elaborate on how this ordering is preserved in a calculation using
solutions of the Schr\"odinger equation as the standard approach to systems beyond three bodies.
To that end, the identity of a NLO-\eftnopi~calculation and the familiar distorted-wave Born approximation (DWBA)
is detailed.
\subsection{\eftnopi~for 2 and 3 nucleons}
As shown in~\cite{vanKolck:1998bw}, the \eftnopi~considers the most general Lagrangian built with nucleon fields
and its derivatives. Initially, terms are ordered by their mass dimension and the tree-level amplitude
is identified with a potential. In momentum space with incoming (outgoing) relative momentum $\vec{p}(\vec{p}\,')$, the latter is given by
\begin{equation}
V(\vec{p},\vec{p}\,')=C_0+C_2\left(\vec{p}\,^2+\vec{p}\,'^2\right)+\ldots\;\;\;.
\end{equation}
\paragraph*{2-body sector}\eftnopi~is an unnatural theory because
in addition to the phenomena defining nuclear physics at the heavy mass scales
$\mn$ (nucleon mass) and $M$ (breakdown scale $\sim m_\pi$), it has to incorporate a rich structure at the small scale
$\aleph\sim\eodim$.
The deuteron binding energy $\eodim$ is small relative to \mpi, and the corresponding
pole of the amplitude can be produced with an unnatural scaling of $C^{(R)}_0\sim 4\pi/(\mn\aleph)$.
In comparison, in a natural theory, the renormalized $(R)$ low-energy constant (LEC) $C_0^{(R)}$
associated with a four-fermi operator is $C_0^{(R)}\sim 4\pi/(\mn M)$.
Due to the unnatural scaling, all iterations (loops) of this interaction are of equal size and need to be considered at the same order
of the EFT expansion.

It has been shown that
$C_2^{(R)}\sim 4\pi/(\mn M\aleph^2)$, \ie, the momentum-dependent interaction scales naturally in contrast
to $C_0$\footnote{See~\cite{Rotureau:2011vf} for a visualization of the different scalings.}.
Each insertion of a $C_2$ vertex is suppressed by a factor of $\aleph/M$ and contributes to
the next higher order in the expansion of the amplitude.

The iteration of the momentum-independent LO interaction and the perturbative insertion of
the NLO vertex is shown without reference to (iso) spin degrees of freedom in
Fig.~\ref{fig_bubble}.
We adopt this power-counting scheme for all $NN$ channels.

In addition to the strong short-range force, electromagnetic interactions contribute
significantly at low energies to the pp amplitude in form of static Coulomb-photon
exchanges.

This exchange implies a different RG parameter dependence,
and an iso-spin-dependent contact interaction has to be introduced to renormalize the pp
channel~\cite{Kong2000137}.

\paragraph*{3-body sector}Until recently (see discussion of~\cite{PhysRevC.89.064003} below),
3-nucleon forces were thought to follow the same pattern in the spin-doublet channel. Namely,
an unnatural enhancement of a momentum-independent 3-nucleon contact interaction due to the
existence of a low-energy scale~\cite{Bedaque:1999ve}, and corrections from 3-nucleon terms
in addition to the 2-nucleon operators which are included perturbatively at an order determined by
dimensional analysis~\cite{Griesshammer:2005ga}. In this framework, the LO potential
\begin{equation}\label{eq_lo-pot}
V_{LO}=C_{0,s}\hat{P}^{(^1S_0)}+C_{0,t}\hat{P}^{(^3S_1)}+D\hat{P}^{(^2S_{1/2})}
\end{equation}
is iterated. It includes a LO 3-body interaction with LEC $D$ in the ${}^2S_{1/2}$ channel.
Predictions are then refined with a perturbative treatment of the NLO potential
\begin{equation}\label{eq_nlo-pot}
V_{NLO}=\left(C_{2,s}\hat{P}^{(^1S_0)}+C_{2,t}\hat{P}^{(^3S_1)}\right)\left(\vec{p}-\vec{p}\,'\right)^2
\end{equation}
which is projected ($\hat{P}^{(^{2S+1}L_J)}$) into decoupled spin-singlet ($s$) and triplet ($t$) $NN$ channels.
As the interaction operates on antisymmetric states, it is sufficient
to consider the squared momentum transfer of the in- and
outgoing nucleon pair, $\vec{q}\equiv\vec{p}-\vec{p}\,'$.
With that LO and NLO, \eftnopi~has been successful in
describing the 3-nucleon system (LO~\cite{Bedaque:1999ve}
and NLO in~\cite{Hammer:2007kq}).
\paragraph*{The Coulomb interaction}
The electromagnetic interaction between two protons is non-perturbative for relative momenta \mbox{$Q\lesssim\alpha\mn\sim 10~$MeV}.
The momentum scale associated with the bound-state pole in the helion channel,
\mbox{$\kty=\sqrt{2\,\mn\eohe/3}\sim 70~$MeV}, is large enough for a perturbative treatment of the Coulomb interaction~\cite{Stetcu:2006ey}.
The perturbative approach would be inappropriate if states with relative pp momentum \mbox{$Q<10~$MeV} contribute significantly
to the helion bound state, as calculations (see~\cite{Wiringa:2013ala} for a recent study) and measurements
(\eg,~\cite{PhysRevLett.49.974}) indicate\footnote{\cite{Wiringa:2013ala,PhysRevLett.49.974} extract occupation numbers. However, it was shown~\cite{Furnstahl:2001xq} that occupation numbers cannot be defined uniquely.
We therefore interpret the large contribution of low-momentum modes of~\cite{Wiringa:2013ala,PhysRevLett.49.974} as
specific to the employed interaction and extraction method.}.


In the following section, we explain how the power counting is implemented in a calculation
of observables by solving the Schr\"odinger equation in coordinate representation.

\subsection{Schr\"odinger formulation of \eftnopi}\label{sec_sgl}
The NLO-\eftnopi~amplitude expressed in terms of wave functions is given by
\begin{eqnarray}\label{eq_wfkt_NLO}
f_{l=0}(p)&=&-\frac{1}{p}\int_0^\infty dr~r~j_0(pr)\left(\mn V_{LO}\right)\psi^{LO}_{p}(r)\nonumber\\
&&-\frac{1}{p^2}\int_0^\infty dr~\left(\mn V_{NLO}\right)\left(\psi^{LO}_{p}(r)\right)^2
\end{eqnarray}
which is known as the distorted-wave Born approximation for the two potentials \mbox{$V_{LO}+V_{NLO}$}
(the derivation and notation is given in Sec.~\ref{sec.app}).

The potentials and associated amplitudes as recapitulated in the previous section need to be renormalized.
In this context, the renormalization scheme comprises two steps \textit{(a), (b)}:

\paragraph{(a)~Regularization of the \eftnopi~potential}
We employ a cutoff scheme convenient for the numerical solution of Eq.~\ref{eq_schroed} in coordinate space:
The renormalized LECs (superscript $\Lambda$) in the 2-body potentials
\begin{equation}\label{eq_vcoord_lo}
V^{\Lambda,2}_{LO}(\vec{r})=\left(C^\Lambda_{0,s}\hat{P}^{(^1S_0)}+C^\Lambda_{0,t}\hat{P}^{(^3S_1)}\right)e^{-\frac{\Lambda^2}{2}\vec{r}^2}
\end{equation}
and
\begin{equation}\label{eq_vcoord_nlo}
V^\Lambda_{NLO}(\vec{r})=\left(C^\Lambda_{2,s}\hat{P}^{(^1S_0)}+C^\Lambda_{2,t}\hat{P}^{(^3S_1)}\right)
\left[\frac{(\Lambda\vec{r})^2-6}{4}\right]e^{-\frac{\Lambda^2}{2}\vec{r}^2}
\end{equation}
depend in contrast to the bare LECs in Eq.~\ref{eq_lo-pot},\ref{eq_nlo-pot} on the regulator. A factor of
$\Lambda^{3\{5\}}/(8\pi^{3/2})$ $(\{N\}LO)$ from the Fourier transform of the momentum space regulator was
lumped into the LECs.

Due to the non-perturbative character of the cutoff dependence of the 3-nucleon amplitude, the promoted 3-nucleon counter term
has to be iterated as well~\cite{PhysRevLett.82.463,Bedaque:1999ve}.
We regularize this term such that its coordinate-space representation reads
\begin{equation}\label{eq_vcoord_lo_tni}
V^{\Lambda,3}_{LO}(\vec{r}_1,\vec{r}_2)=\left(\frac{\Lambda^3}{8\pi^3/2}\right)^2\cdot D^\Lambda~e^{-\frac{\Lambda^2}{2}\left(\vec{r}_1^2+\vec{r}_2^2\right)}\;\;\;
\end{equation}
and adopt the power counting of~\cite{Bedaque:2002yg}~in which this is the only 3-body operator up to NLO in the triton channel.
A LO-\eftnopi~calculation in the 3-body sector then proceeds by solving Eq.~\ref{eq_schroed} with the sum of Eq.~\ref{eq_vcoord_lo} and Eq.~\ref{eq_vcoord_lo_tni}.

\begin{figure}[tb]\begin{center}
\includegraphics[width=1. \textwidth]{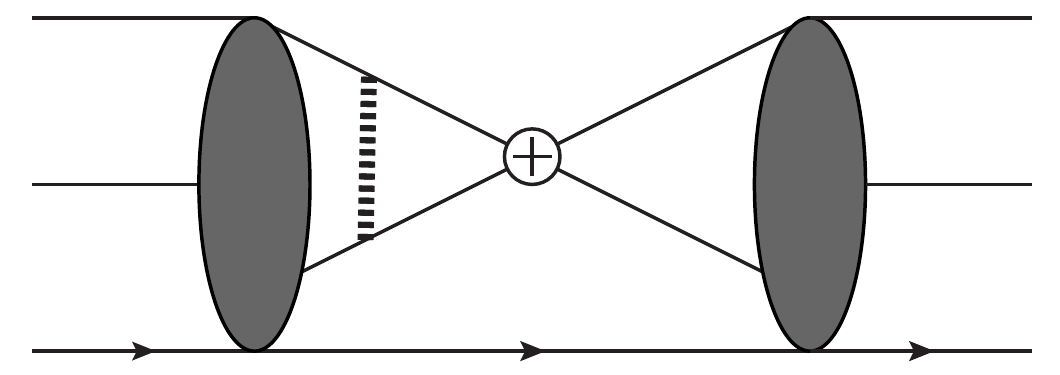}
\caption{\label{fig_triton} Graphical representation of the lowest order Coulomb correction of the helion binding energy.
Triton wave functions are indicated by gray-filled blobs with three incoming lines. An (no) arrow shows the forward time direction for
neutrons (protons), the exchanged Coulomb photon dashed, and the pp vertex is the circle cross.}
\end{center}
\end{figure}

Based on the discussion above about typical momenta in the 3-nucleon bound system, we consider the Coulomb interaction as a perturbation \textit{and} as a non-perturbative effect.
As a perturbation, the Coulomb potential 
\begin{equation}\label{eq_nlo_coul}
V^\Lambda_{C}(\vec{r})=C^\Lambda_{pp}\hat{P}_{pp}^{(^1S_0)}e^{-\frac{\Lambda^2}{2}\vec{r}^2}+\frac{e^2}{4|r|}\;\;\;.
\end{equation}
contributes in the same way as the NLO-\eftnopi~interaction in Eq.~\ref{eq_vcoord_nlo} to the 3-body amplitude (see cartoon in Fig.~\ref{fig_triton}).
Through the projector $\hat{P}_{pp}^{(^1S_0)}$, the LEC $C^\Lambda_{pp}$ acts between protons, only.
The proper non-perturbative treatment defines a new LO potential as the sum 
of Eq.~\ref{eq_vcoord_lo} and Eq.~\ref{eq_nlo_coul}.

For the cutoff, we chose a range from $660~$MeV to $2.4~$GeV to cover an interval in which the results
of~\cite{PhysRevC.89.064003} display significant cutoff dependence. As we employ a variational method to
solve the 3-body problem, reaching convergence in excited states is numerically challenging. Thus we limit
the analysis to this range where we can find one bound state below and two above a critical cutoff $\Lambda_\text{crit.}\sim 1230~$MeV.
\paragraph{(b)~Calibration of LECs}
We calibrate/renormalize the LECs such that the amplitudes Eq.~\ref{eq_wfkt} (LO) and Eq.~\ref{eq_wfkt_NLO} (NLO) resemble appropriate data.
Scattering phase shifts ($\delta$) at low energy are the canonical choice.
To express the corresponding amplitude in terms of effective-range parameters,
two steps are involved: expand $p\cot\delta(p)$ about some $p$ (effective-range expansion);
expand the ensuing amplitude, $f(p)=\left(p\cot\delta-ip\right)^{-1}$, in a power series consistently, \ie,
about a momentum the EFT expansion is expected to converge rapidly, and up to the same power as the EFT.
\par
At LO, we expand about $|\vec{p}\,|=0$, and thus have to adjust the LECs such that the amplitude
from Eq.~\ref{eq_wfkt} matches the scattering length ($a$) of the respective channel, \mbox{$\lim_{p\to 0}f(p)=-a$}.
For the NLO calculation, it is prudent to expand about a momentum ($p_0$) which is typical for the system
under investigation and \textit{not} about zero\footnote{See Bethe's original ERE expansion about the deuteron
pole~\cite{PhysRev.76.38} or
the fixing of the residue in~\cite{Phillips:1999hh} as examples of wise choices for the momenta to expand about.}.
The three-nucleon system contains the singlet and triplet
states as subsystems and momentum scales in those systems are given by the poles of the respective amplitude in
the complex momentum plane:
\begin{equation}\label{eq.poles}
\gamma=\pm i\,r^{-1}\left(1-\sqrt{1-2ra^{-1}}\right)=\mp i\scalebox{0.9}{$7.88\;(S=0)\atop 45.7\;(S=1)$}~\text{MeV}\;\;.
\end{equation}
The logical choice is thus: $p_0=|\gamma|$ because we use positive
energy scattering states as asymptotic boundary conditions to obtain amplitudes from 
Eq.~\ref{eq_wfkt},\ref{eq_wfkt_NLO}).
Explicitly, we match Eq.~\ref{eq_wfkt_NLO} to the \textit{Taylor expansion} (we spare the reader the
lengthy expression) of
\begin{equation}\label{eq_ere_nlo}
f(p)=\left(-|\gamma|+\frac{r_0}{2}(\gamma^2+p^2)-ip\right)^{-1}
\end{equation}
about the pole $p_0=|\gamma|$ in the $NN$ singlet and triplet channel, respectively.

The LEC $C^\Lambda_{pp}$ is renormalized to the pp scattering length, $\app=-7.81~$fm, by matching the pp amplitude to
the generalized effective range expansion~\cite{PhysRevC.26.2381} for charged particles. For this
calibration, the interaction in Eq.~\ref{eq_nlo_coul} is iterated because $\app$ is extracted from the amplitude at
momentum $\ll 1~$MeV.
Whether or not the thus obtained $C^\Lambda_{pp}$ does renormalize the 3-nucleon amplitudes,
if considered perturbatively, as detailed below Eq.~\ref{eq_nlo_coul}, validates or invalidates the power counting.
We rely on a numerical calculation for this assessment.
The renormalization scheme as defined above defies an analytic argument~\footnote{As one presumably given in~\cite{Konig:2015aka},
where the authors address the same complex as this article does, and of which we became aware while finalizing this manuscript.}
because, first, the bare Coulomb wave functions are approximated and separately cut off at small distances, and second, because we
do not regularize dimensionally but with a cutoff function.

\paragraph{Practical implementation of \eftnopi}
The LO LECs, $C_{0,s/pp}^\Lambda,C_{0,t}^\Lambda$, are calibrated by solving the Schr\"odinger equation~\ref{eq_schroed}
in coordinate representation and a matching of the corresponding amplitude Eq.~\ref{eq_wfkt} to the respective
scattering length. For the 3-nucleon LEC, the thus obtained 2-body LECs are held fixed, while
$D^\Lambda_{(*)}$ is adjusted such that Eq.~\ref{eq_schroed} has a negative energy eigenvalue equal to $\eotri$
in the triton channel corresponding to the ground or first excited state (indicated by a $*$ and explained at the end of this section).
We predict solely $\eohe$ with this theory for a non-perturbative Coulomb interaction.
If it is treated in perturbation theory, $\eohe=\eotri$ at LO.

At NLO, too, we first renormalize the 2-body sector. First, we minimize the difference between the DWBA amplitude Eq.~\ref{eq_wfkt_NLO} and the
power expansion of Eq.~\ref{eq_ere_nlo}. This minimization varies the LECs, $C_{0/2,s}^\Lambda,C_{0/2,t}^\Lambda$, from which
the unperturbed wave functions are calculated as in LO above, until a termination condition is satisfied.
The optimization starts with the LO values for $C_{0,s/t}^\Lambda$,
and $C_{2,s/t}^\Lambda=0$. The implied adjustment of LECs of momentum-independent vertices has an analog in the
integral approach (see \eg,~\cite{vanKolck:1998bw}).
For the calibration of $D^\Lambda_{(*)}$, the 2-body LECs are fixed. With $C_{0,s/t}^\Lambda$, which
\textit{differ} from the LO values, and a trial $D^\Lambda_{(*)}$ a ground (excited) state wave function is
calculated via Eq.~\ref{eq_schroed}. The energy eigenvalue of that state is shifted by the matrix element of that state and
the NLO potential in Eq.~\ref{eq_vcoord_nlo}. $D^\Lambda_{(*)}$ is optimized such that this sum equals the triton binding energy.

Thereby, renormalization constraints of NLO-\eftnopi~used in this work incorporate 6 data points:
the neutron-proton scattering lengths and effective ranges for the singlet and triplet
channel, the proton-proton scattering length, and the triton binding energy. The pp and np-singlet effective ranges
were assumed to be identical, and hence the NLO interaction in Eq.~\ref{eq_vcoord_nlo} does not break iso-spin symmetry.

\begin{figure}[tb]\begin{center}
\includegraphics[width=1. \textwidth]{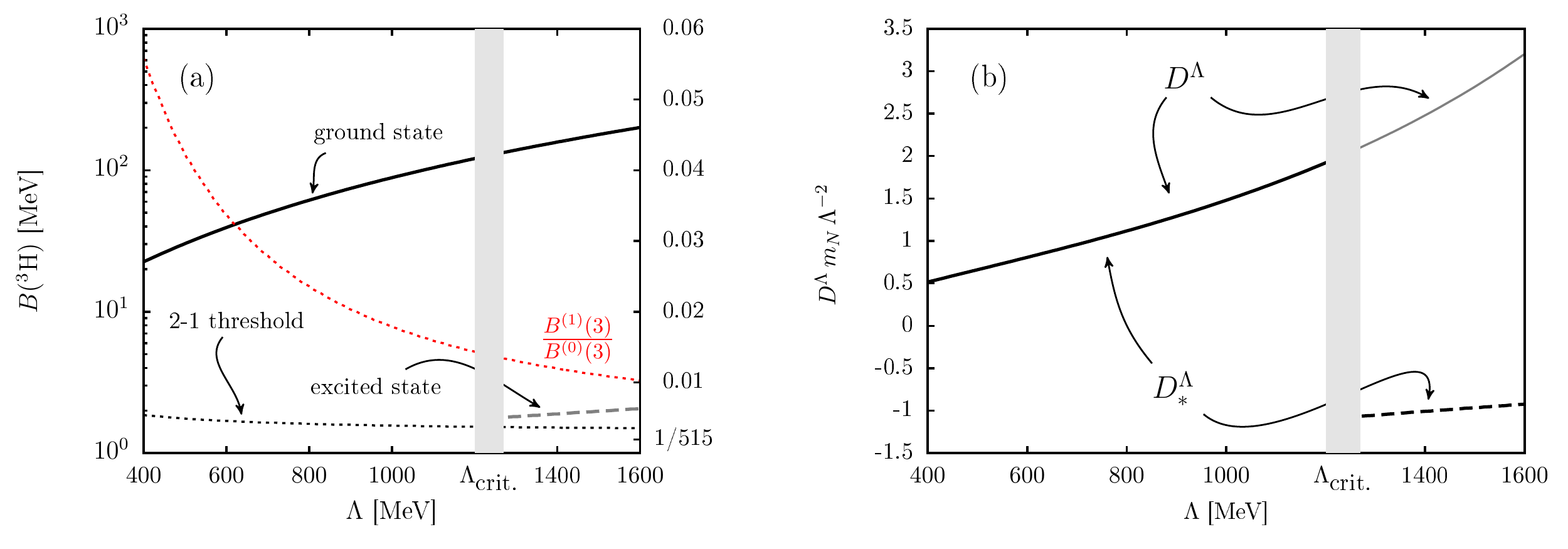}
\caption{\label{fig_limit-cyc} Cutoff dependence of the ground (black solid)
and first-excited-state (gray dashed) binding energies (left MeV Y-scale ),
and their ratio (dotted red, right Y scale) for $D^\Lambda_{(*)}=0$ (panel (a)).
In panel (b), the running of the dimensionless 3-nucleon
LEC with the cutoff is shown. For $\Lambda\lesssim\Lambda_\textrm{crit.}$, only the
ground state exists below the neutron-deuteron breakup
threshold (thin dotted line (a)) with a corresponding
repulsive 3-body LEC (solid black line (b)).
For $\Lambda\gtrsim\Lambda_\textrm{crit.}$, an excited state appears,
which can be matched to $B^{(1)}(3)\sim 8.5~$MeV with an attractive
3-body LEC (black dashed (b)), while the deeply-bound
ground state requires an increasingly repulsive LEC (thin gray line (b)) to meet $B^{(0)}(3)\sim 8.5~$MeV.
}
\end{center}
\end{figure}

Noteworthy are the two renormalization conditions we employed
for the 3-nucleon interaction.
First, we identified the \textit{shallowest} state
of the nnp-spin-doublet system with the triton.
This state can be the ground or an excited state, if the
EFT tenet is adopted that effects of deep states, \ie, composed predominantly
of modes with momentum beyond the breakdown scale,
on predictions of a renormalized EFT can be removed order by order.
In a scenario where the 3-nucleon spectrum \textit{without} \mbox{3-body} force sustains 
two bound states (results in Fig.~\ref{fig_limit-cyc}~(a) for $\Lambda>\Lambda_\textrm{crit.}$ or
, \eg, Fig.~6 in~\cite{Bedaque:1999ve}), the adjustment of the 3-body parameter to
fix the excited state to the triton will leave the deep state in the spectrum.
The corresponding LEC $D^\Lambda_{*}$ exhibits a discontinuous jump between 
two branches: a repulsive branch
(solid black in Fig.~\ref{fig_limit-cyc}~(b)) which matches
a deeply-bound ground state to the triton, and an attractive branch
(dashed black Fig.~\ref{fig_limit-cyc}~(b)) which lowers an excited state
entering at $\sim\Lambda_\textrm{crit.}$ to the triton
(see also the discussion of Figs.~\ref{fig_3He_pert}~and~\ref{fig_3He}, below).
The two branches do not resemble the log-periodic cutoff dependence of
the dimensionless 3-body parameter found analytically with auxiliary dibaryon
fields~\cite{PhysRevLett.82.463}. We assume that the running of our 3-body parameter
continues periodically, too. While the detailed functional dependence of the cycle is regulator
dependent, a constant ratio between the two shallowest 3-nucleon states is widely believed to be a
universal feature of nuclei; all regulators must approximate the same ratio barring higher-order
effects. The second assumption about the regulator employed here is thus: the second excited bound state
which enters, considering the small slope of $D^\Lambda_*$ at $\Lambda\gg 2.4~$GeV,
has a binding energy which is relative to the first excited state
in the same universal ratio as extrapolated from the calculated spectrum.
From the red-dotted line in Fig.~\ref{fig_limit-cyc}~(a), we find $B^{(1)}(3)/B^{(0)}(3)$ 
to converge to a value close to the universal ratio of $\sim 1/515$ which was derived for
$a\to\pm\infty$~\cite{1970PhLB...33..563E}.

The \textit{second} condition allows only one bound state in the spectrum and adjusts $D^\Lambda$ to fix this state's
energy to $\eotri$. This LEC will not follow a limit cycle because even beyond
the critical cutoff it adjusts the deep state and \textit{not}
the incoming shallow one.
Hence, above some cutoff\footnote{Here, this value is apparently $<400~$MeV.},
this procedure will always produce
a repulsive 3-body force (see thin gray line as
continuation of solid black in Fig.~\ref{fig_limit-cyc}~(b)).

\paragraph{Toolbox}
Optimization problems are solved with the \textit{BFGS} quasi-Newton method~\cite{nocedal1999numerical} as
implemented in the \textit{numPy}-$1.8.2$ library.
To obtain 2-body scattering wave functions, we alternated between three methods to minimize the possibility
of a mistake in the numerical implementation of either method: a \textit{Numerov} integration, a version~\cite{0953-4075-36-19-013}
of the \textit{variable phase} method, and a \textit{variational} method
with a Gaussian trial wave function.
Three-body wave functions and matrix elements were
calculated with the \textit{refined-resonating-group method}
(\rgm, original formulation:~\cite{Wheeler:1937zza,Wheeler:1937zz}; refinement and triton/$^3$He
implementation:~\cite{hmh-rrgm,Kirscher:2011zn})
\footnote{As the \rgm~is readily replaced with another method, all
necessary techniques are standard and the described implementation can easily be 
reproduced.}. The variational \rgm, as applied here,
is significantly less accurate than the Numerov integration and
the variable-phase method. We estimated its dominant numerical uncertainty by a rescaling of the
variational Gaussian width parameters. With fixed LECs, the variation in $\eohe$, induced by this
multiplication of the widths on the two Jacobi coordinates by factors $\in(0.2,2)$, was found to
be $\lesssim0.1~$MeV.
\section{Helium-3 Results}
\begin{figure}[tb]
  \centering
\includegraphics[width=1.0 \textwidth]{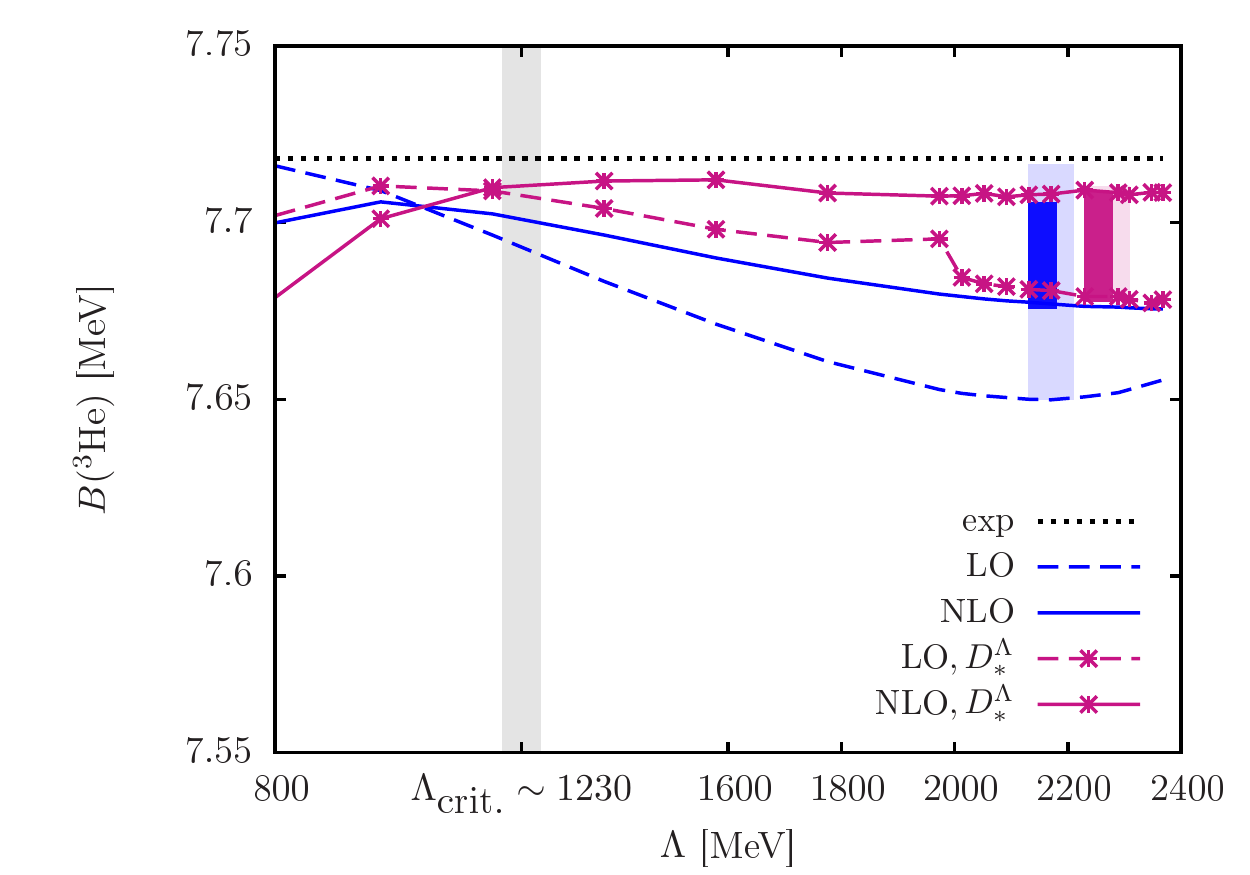}
\caption{\label{fig_3He} The Helium-3 binding energy calculated from \eftnopi~with a \textit{non-perturbative} Coulomb interaction.
(NLO)LO results are shown as (solid)dashed lines and were obtained with the \rgm. Results admitting deep 3-body state in the spectrum, \ie, Helium-3 is an excited state, are marked with stars.
The assessed $\Lambda$ uncertainty is indicated by the height of the transparent (LO) and opaque (NLO) rectangle.
An additional 3-body bound state enters the spectrum (gray band)
at $\Lambda_\text{crit.}\sim 1230~$MeV.}
\end{figure}

\begin{figure}[tb]
  \centering
\includegraphics[width=1.0 \textwidth]{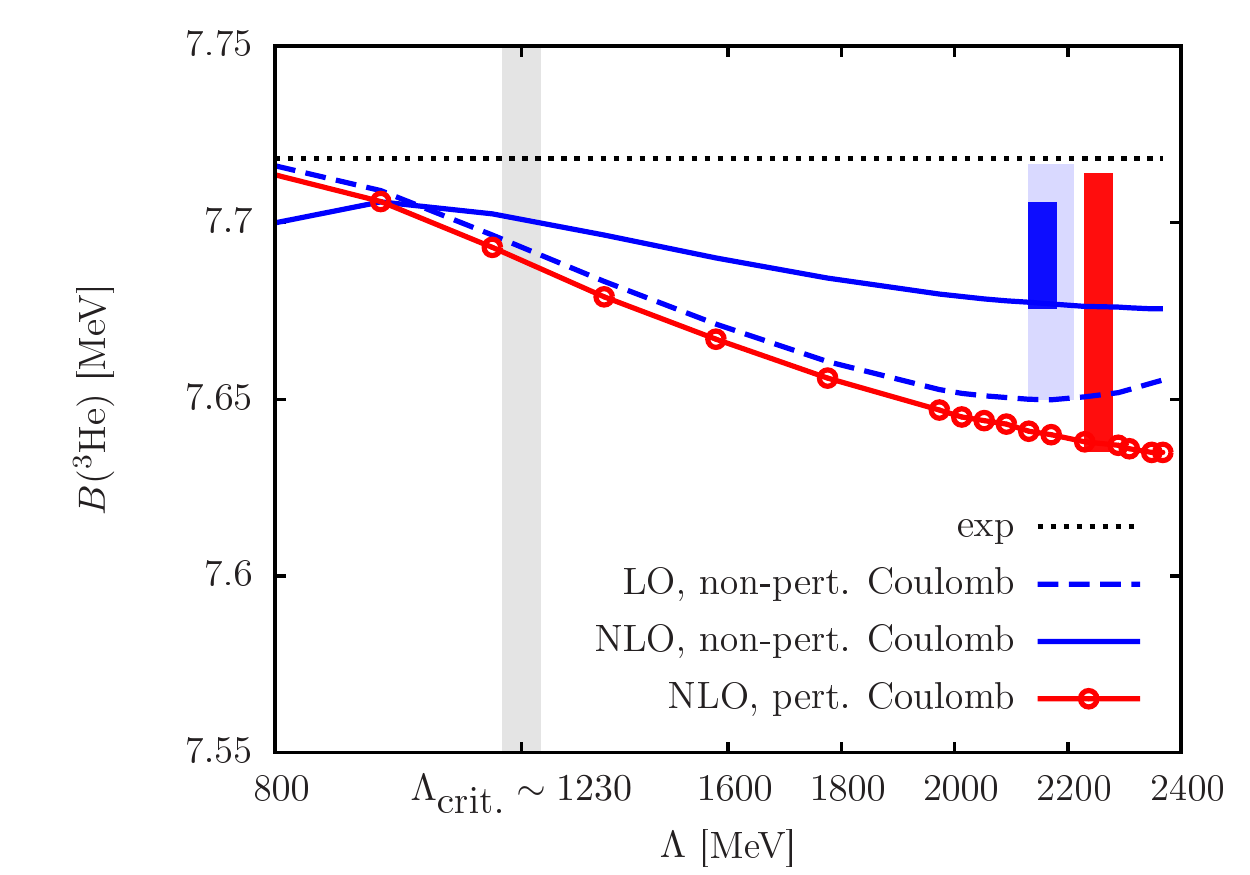}
\caption{\label{fig_3He_pert} Comparison between \textit{perturbative} (circles) and \textit{non-perturbative} (no marker) Coulomb treatment in Helium-3 with
\eftnopi. (NLO)LO results are shown as (solid)dashed lines.
The horizontal dotted line represents the experimental $^3$He binding energy.
Transparent (LO) and opaque (NLO) rectangle heights indicate $\Lambda$ uncertainty. At \mbox{$\Lambda_\text{crit.}\sim 1230~$MeV} (gray band),
an additional 3-body bound state enters the spectrum.}
\end{figure}

The \eftnopi~as parameterized in the previous section was employed at LO and NLO to
postdict the helion binding energy. Results are presented in Figs.~\ref{fig_3He}~and~\ref{fig_3He_pert}~as functions
of the cutoff to assess the sensitivity of the result to the renormalization scheme. This error analysis
constitutes the canonical, \textit{a posteriori} justification
for the power counting of a cutoff EFT.

First, results with the \textit{non-perturbative} Coulomb interaction are shown in Fig.~\ref{fig_3He}.
For both 3-body renormalization conditions, with (starry lines) and without the deep trimer,
NLO results (solid lines) are found more stable for $\Lambda\to 2.4~$GeV than at LO (dashed) in the same limit.
The total variation of $\eohe$~over the considered cutoff range is highlighted in
Figs.~\ref{fig_3He}~and~\ref{fig_3He_pert}~by the height of overlapping rectangles.
Comparing these uncertainties in $\eohe$, we find the NLO results
without a deep trimer about twice as accurate at
NLO (blue, opaque rectangle at $\sim 2.2~$GeV) relative to
LO (blue, transparent overlapping rectangle).
With a deep trimer, LO and NLO results are of about the same
accuracy (red, overlapping opaque and transparent rectangles at $\sim 2.3~$GeV).
The significant difference in LO uncertainties,
depending on the chosen 3-body renormalization, with or without
a deep state, exemplifies the need to test the
renormalization-scheme independence beyond a
cutoff-parameter variation with otherwise fixed regulator
shape and identical matching conditions, \ie, data input.
A comprehensive analysis (see~\cite{Furnstahl:2014xsa}~for
a recipe and~\cite{Epelbaum:2014sza}~for its application),
which would thus vary matching conditions and regulator shape, is
required to assess the convergence rate of the EFT. Here, we aim to
verify the consistency of the EFT's power counting, and it suffices to demonstrate convergence
of predictions in a limit which removes the arbitrary regulator. To that end, we observe that
the stability of all four curves for $\Lambda\to 2.4~$GeV does not indicate
a failure of the power-counting of \eftnopi~up to NLO with a non-perturbative
long-range Coulomb interaction. 

If the Coulomb interaction is counted as a \textit{perturbation}, the resultant postdiction for $\eohe$~is shown
in Fig.~\ref{fig_3He_pert} (red line with circles). The LO result, in this counting scheme, coincides with
the triton binding energy and is not shown as it is cutoff independent by construction.
Treating the Coulomb interaction at the same order as the most significant
momentum-dependent terms of \eftnopi~and
\textit{not} at LO, where the kinetic energy operator
$\vec{\nabla}^2/(2\mn)$ and the 4- and 6-fermi momentum-independent
terms are in balance to yield the shallow deuteron and
triton states, is supported further by a comparison between the
matrix elements (Eq.~\ref{eq_dwba_me}) of the Coulomb (Eq.~\ref{eq_nlo_coul})
and NLO-\eftnopi~(Eq.~\ref{eq_vcoord_nlo}) operators between LO triton wave functions.
Both contributions, Coulomb and NLO to $\eohe$~were
found to be of $\mathcal{O}(1)$ in the considered cutoff interval
while, in contrast, the kinetic energy and LO operators assume
for $\Lambda>1~$GeV values of $\mathcal{O}(10^2)$.
In fact, perturbative Coulomb and NLO operators yield results close to the
non-perturbative LO values
(compare encircled solid and dashed lines in Fig.~\ref{fig_3He_pert}).
This proximity is consistent with the key finding of this article, namely the perturbative
character of the Coulomb interaction in ${}^3$He. The iteration of the Coulomb
exchange should then produce only higher-order-in-$\alpha\mn/Q$ corrections, which
is precisely what we observe.
Behavior like the increased splitting between the encircled
solid perturbative NLO line and the dashed LO line in
Fig.~\ref{fig_3He_pert} for $\Lambda\gtrsim 2~$GeV is well within the limits of the numerical
accuracy ($\pm0.1~$MeV). Instabilities of that magnitude and shape
(see also the step-like behavior of the star-dashed line in Fig.~\ref{fig_3He})
are thus unlikely to resemble divergent components of the amplitude.

Assuming that the numerical uncertainty affects all results equally, shifting them
by a similar amount if the variational parameters are refined,
the dependence on $\Lambda$ at NLO is found to be of the same
magnitude as the LO uncertainty for non-perturbative Coulomb
(compare transparent box height on the left with opaque height of the right box).
Furthermore, the convergence with $\Lambda$ is slower
and has not yet reached a plateau at $\Lambda\sim 2.4~$GeV as in Fig.~\ref{fig_3He}.
The uncertainty represented by the
height of the red opaque rectangle is thus nothing but a lower bound.
The observed cutoff dependence is, however, not
indicative for a failure of the power counting since
the assessed lower bound of the uncertainty of $\sim 0.1~$MeV is small compared with
the 3-nucleon limit cycle~\cite{PhysRevLett.82.463} for $\eotri$,
or the $\sim 1.8~$MeV change in $\eohe$ found in the NLO
analysis of Helium-3~\cite{PhysRevC.89.064003} over the same cutoff range we cover.

Given the different regularization schemes and therefore RG trajectories followed
by the respective $\Lambda$
in~\cite{PhysRevC.89.064003} and this work, a comparison of the
ranges in which $\Lambda$'s were varied is
non-trivial.
To assess whether or not sensitivity to short-distance structure is probed to the
same extent with two arbitrary regularization schemes, comparing the ranges over which the
respective parameters are varied is not productive. Here, we select an observable and compare
cutoff ranges based on variations in that observable. Regulator variations are considered
similar if they induce variations of the same order of magnitude in the observable.
For this observable, we chose the universal ratio of consecutive bound
states in the 3-body system. We showed above (red-dotted line in Fig.~\ref{fig_limit-cyc}~(a))
that our result for the binding-energy ratio of excited and ground state is almost converged
and little variation is expected at higher cutoffs. Further cutoff variation will not
increase the uncertainty in this observable and thus would not yield further
insight to the renormalization-scheme dependence. In this metric, our cutoff range is comparable
to the $\sim10^5~$MeV-wide interval analyzed, \eg, in~\cite{PhysRevLett.82.463}.

In more detail, shallow 3-body states are created
at threshold if no 3-nucleon force is adjusted and
the two (in~\cite{PhysRevC.89.064003} the three) body regulator is varied.
In the scheme employed here, the first of such states enters at a critical cutoff
$\Lambda_\text{crit.}\sim 1230~$MeV (gray band in Figs.~\ref{fig_3He} and~\ref{fig_3He_pert}).
In~\cite{PhysRevC.89.064003}, the first divergence of the limit cycle of the LO 3-body LEC is located at a cutoff of $\sim 1.8~$GeV.
The next bound state enters the spectrum at $\sim 36~$GeV in~\cite{PhysRevC.89.064003}. For our method it is impractical to
probe $\Lambda$ values that high. Nevertheless, assuming our cutoff scheme would require a similar counter term as
the one promoted in~\cite{PhysRevC.89.064003}, the cutoff range between 0.6~MeV and 2.4~GeV assessed here should suffice.
In this range, the NLO 3-body LEC of~\cite{PhysRevC.89.064003} completes almost a complete cycle including
zero. As the frequencies of the LO and NLO limit cycles (compare Figs.~6 and 9 in~\cite{PhysRevC.89.064003}) are similar
and the first LO divergence occurs almost at the same cutoff value, we na\"ively expect a similar $\eohe$ variation
as in~\cite{PhysRevC.89.064003}.
However, the change we observe in the helion binding energy within this cutoff range is an order of magnitude smaller~---~note the 
y-scale in Figs.~\ref{fig_3He} and~\ref{fig_3He_pert} spans only $200~$keV~---~and thereby
consistent within the EFT and numerical uncertainty. 

To conclude, a power counting with LO comprised by the kinetic, three 4-fermi ($pp$, $np$ singlet/triplet)
and one 6-fermi operator treated non-perturbatively, and NLO adding perturbatively two momentum-dependent 4-fermi interactions is valid for the description of the 2 and 3-nucleon bound states.
This counting is useful regardless of including the Coulomb interaction perturbatively or non-perturbatively.
We argued that the seeming difference to the work of~\cite{PhysRevC.89.064003}, in the case of a non-perturbative Coulomb force, is
related to the regularization scheme which has a significant impact on the power counting.
\newpage
\section{Conclusion}
The renormalization-group dependence of the helion ground state was analyzed with the \eftnopi.
The binding energy of that state was found invariant with respect to an RG parameter which
rescales the 2- and 3-body interaction in a similar fashion. The residual uncertainty
in this next-to-leading-order analysis was found insensitive to the iteration of the exchange of Coulomb
photons, thus justifying their perturbative treatment.

The result seemingly contrasts an analysis with a non-perturbative Coulomb interaction formulating the few-body
problem with auxiliary fields as a set of Lippmann-Schwinger equations. As both formulations are identical
except for the regularization and the RG flow thereby parameterized, we stress that despite the suggestive $\Lambda$ nomenclature
common to both calculations, the methods differ significantly in that procedure.
The dimensional regulator for the 2-body combined with the cutoff regulator in the 3-body propagator follow a critical trajectory while the
regularization chosen in this work does not. This difference between combining dimensional regularization with power
divergence subtraction (PDS) with a cutoff regulator~\cite{PhysRevC.89.064003},
on the one hand, to a cutoff-regulated theory for both the 2- and
3-nucleon amplitudes on the other hand is an unsolved problem.
Its importance reaches beyond \eftnopi, considering
the analytical and numerical efforts to match chiral effective theories with \eftnopi~to determine the quark-mass
dependence of the nuclear force rigorously (\cite{Epelbaum:2002gb} and~\cite{Hammer:2007kq} are EFT-inspired models in that direction):
chiral interactions for few-body systems are completely cutoff
regulated while \eftnopi, barring~\cite{Kirscher:2009aj,Stetcu:2006ey,Platter:2004zs},
implementations mix cutoff with dimensional-PDS regularization.
The discrepancy uncovered here, shows that the two schemes differ in the presence of a long-range
interaction, namely the Coulomb force, and thus asks for a generalization of the analysis of~\cite{Birse1999169}
to the three-body sector.

The above methodology of solving \eftnopi~in coordinate space with a regulated 2- and 3-body potential
was chosen for its practicality to analyze larger nuclei. Regardless of the applicability of \eftnopi~in
the 4-nucleon system and beyond, the larger momentum scale in those nuclei suggests that a perturbative
treatment of the Coulomb interaction for their ground states converges even faster than in helion.
Even for Helium-3 and the triton, our results do not rely on \eftnopi~as the short-distance interaction.
A perturbative Coulomb force will be useful for any nuclear interaction describing consistently the 2-nucleon
scattering lengths and effective ranges, and the triton ground state.

Considering the ongoing struggle in finding a RG invariant formulation of chiral perturbation theory for few-nucleon
systems~\cite{Nogga:2005hy,Long:2011xw,Long:2012ve,Valderrama:2011mv,Valderrama:2009ei}, we
conclude with noting the intriguing application of the above treatment of photons, \ie, iterated and
renormalized in the 2-body sector and as a perturbation in larger systems, to
pions\footnote{\cite{Baru:2015ira} is a recent example for reinterpreting a methodology originally
devised for long-range Coulomb and short-range nuclear interactions through an identification of the pion-exchange
as long ranged.}.
\appendix
\section{The distorted-wave-Born amplitude}\label{sec.app}
We recapitulate the distorted-wave Born approximation of a scattering amplitude for the sum of two potentials:
the short-range strong interaction, for which the Schr\"odinger equation is solved, and the long-range Coulomb
interaction augmented with short-range operators, which are then considered as a perturbation.
We follow the standard of Taylor~\cite{Taylor1972}.

\begin{figure}[tb]\begin{center}
\includegraphics[width=1. \textwidth]{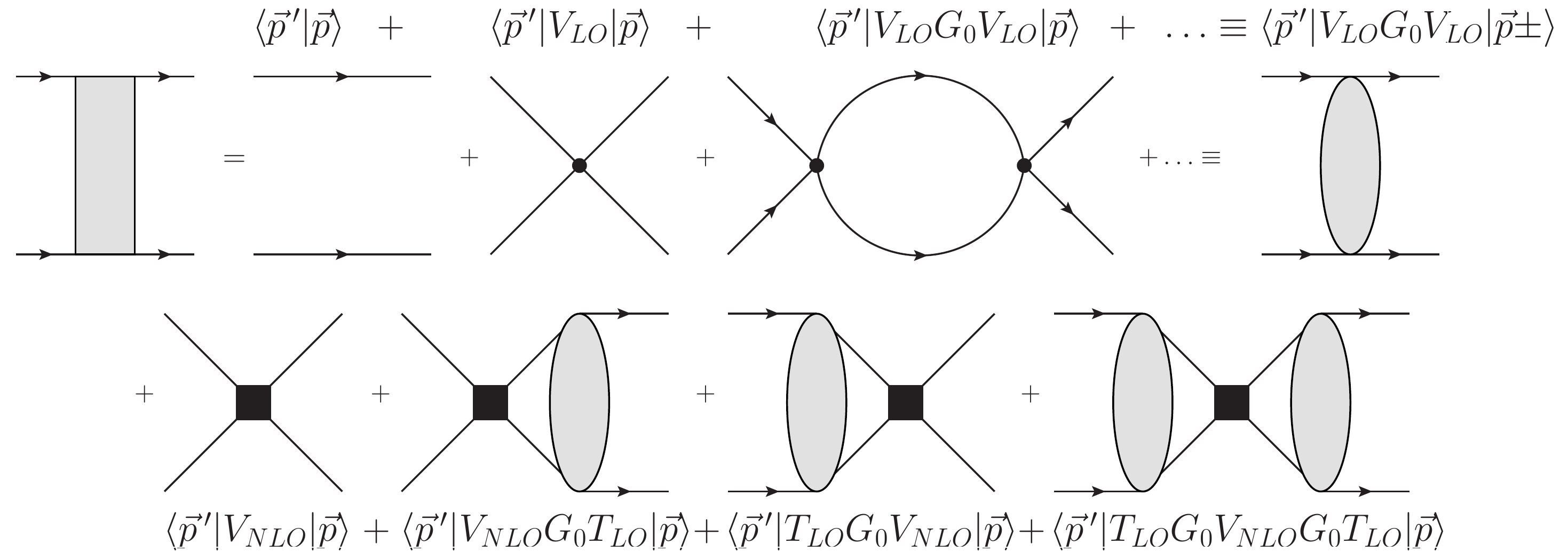}
\caption{\label{fig_bubble} Diagrammatic representation of the NLO-\eftnopi~amplitude (gray, filled rectangle) as given in Eq.~\ref{eq_wfkt_NLO}. The LO amplitude (gray, filled blob) is defined in the first line with free in and outgoing states with relative momenta $\vec{p}$ and $\vec{p}\,'$. The
perturbative insertion of the NLO vertex (square, Eq.\ref{eq_nlo-pot}) is shown in the second line.}
\end{center}
\end{figure}

The full scattering states $|\vec{p}\pm\rangle_V$ corresponding to an interaction $V$
and in/outgoing boundary conditions ($\pm$)
obey the stationary Schr\"odinger \textit{differential} equation
\begin{equation}\label{eq_schroed}
\left(H_0+V\right)|\vec{p}\pm\rangle_V=E_p|\vec{p}\pm\rangle_V\;\;.
\end{equation}
The wave function $\psi^V_p(\vec{r})=\langle\vec{r}\,|\vec{p}\pm\rangle_V$
with \mbox{$E_p=p^2/\mn$} can also be obtained from the Lippmann-Schwinger \textit{integral} equation
\begin{equation}\label{eq_LSS}
\langle\vec{r}\,|\vec{p}\pm\rangle_V=\langle\vec{r}\,|\vec{p}\,\rangle_0+\langle\vec{r}\,|G_0(E_p\pm i\epsilon)V|\vec{p}\pm\rangle_V\;\;\;.
\end{equation}
A free state $|\vec{p}\,\rangle_0$ and the Green's operator $G_0$ follow from Eq.~\ref{eq_schroed} with $V=0$.
The $T$ operator acts
in the space of free states and is introduced as $T(E_p\pm i\epsilon)|\vec{p}\,\rangle_0=V|\vec{p}\pm\rangle_V$.
In terms of $T$, Eq.~\ref{eq_LSS} reads
\begin{eqnarray}\label{eq_LST}
{}_0\langle\vec{p}\,'|T|\vec{p}\,\rangle_0&=&{}_0\langle\vec{p}\,'|V|\vec{p}\,\rangle_0+
{}_0\langle\vec{p}\,'|VG_0T|\vec{p}\,\rangle_0\nonumber\\
&=&{}_0\langle\vec{p}\,'|V|\vec{p}\,\rangle_0+{}_0\langle\vec{p}\,'|VG_0V+VG_0VG_0V|\vec{p}\,\rangle_0\nonumber\\
&&+\ldots\;,
\end{eqnarray}
where the iterative solution in the second line is graphically represented by the diagrams in
the first line of Fig.~\ref{fig_bubble} if $V= V_{LO}$.
The full amplitude calculated via
\begin{eqnarray}\label{eq_wfkt}
f_{l}(p)&=&-\mn\int_0^\infty drr^2{}_0\langle\vec{p}\,'|\vec{r}\rangle V\langle\vec{r}\,|\vec{p}\pm\rangle_{LO}\nonumber\\
&\stackrel{l=0}{=}&-\frac{1}{p}\int_0^\infty dr~r~j_0(pr)\left(\mn V\right)\psi^{LO}_{p}(r)
\end{eqnarray}
is thus identical to the one obtained from
\begin{equation}\label{eq_t}
f_{l=0}(p)={}_0\langle\vec{p}\,'|T(E_p\pm i\epsilon)|\vec{p}\,\rangle_{0}\;\;\;.
\end{equation}
In the \eftnopi~power counting, a coupling to partial waves with $l>0$ becomes relevant beyond NLO.
With \mbox{$p\equiv(\mn E_p)^{1/2}$}, spherical Bessel function, $j_0$, and
the radial wave function solving \mbox{$(\partial_r^2-\mn V+p^2)\psi_p(r)=0$} we follow the notation
of~\cite{Taylor1972}.
The LO-\eftnopi~amplitude is thus either calculated via Eq.~\ref{eq_wfkt} with a wave function 
which is the solution of a differential equation, or as the solution of an integral equation
(Eq.~\ref{eq_LST} and Fig.~\ref{fig_bubble} first line).
However, to solve Eq.~\ref{eq_LST}, knowledge of the free wave functions $\langle\vec{r}\,|\vec{p}\,\rangle_0$
is needed.

Reformulating Eq.~\ref{eq_wfkt} in a basis given by the full LO solutions,
$\langle\vec{r}\,|\vec{p}\pm\rangle_{LO}$, the Born approximation yields the NLO-\eftnopi~amplitude as
shown in Fig.~\ref{fig_bubble}. To see this, one identifies $H_0\equiv T_{\tiny\text{kin}}+V_{LO}$ and
$V\equiv V_{NLO}$. Now, assuming
\mbox{$|\vec{p}\pm\rangle_{LO}=|\vec{p}\,\rangle_0+G_0(E_p\pm i\epsilon)V_{LO}|\vec{p}\pm\rangle_{LO}$}
can be solved, Eq.~\ref{eq_LST} for the full (\textit{not} yet NLO) amplitude is
\begin{equation}\label{eq_LST_NLO}
{}_0\langle\vec{p}\,'|T|\vec{p}\,\rangle_0={}_{LO}\langle\vec{p}\,'\text{-}|V_{LO}|\vec{p}\,\rangle_0+{}_{LO}\langle\vec{p}\,'\text{-}|V_{NLO}|\vec{p}+\rangle_{NLO}\;\;.
\end{equation}
Approximating the full solution
$$|\vec{p}+\rangle_{NLO}\approx|\vec{p}+\rangle_{LO}=|\vec{p}\,\rangle_0+G_0T_{LO}|\vec{p}\,\rangle_0$$
yields for the second term in Eq.~\ref{eq_LST_NLO}:
\begin{eqnarray}\label{eq_dwba_me}
{}_{LO}\langle\vec{p}\,'\text{-}|V_{NLO}|\vec{p}+\rangle_{NLO}&\approx&
{}_0\langle\vec{p}\,'|V_{NLO}|\vec{p}\,\rangle_0
+{}_0\langle\vec{p}\,'|T_{LO}G_0 V_{NLO}|\vec{p}\,\rangle_0
\nonumber\\&&+{}_0\langle\vec{p}\,'|V_{NLO}G_0T_{LO}|\vec{p}\,\rangle_0
+{}_0\langle\vec{p}\,'|T_{LO}G_0 V_{NLO}G_0T_{LO}|\vec{p}\,\rangle_0\nonumber\\
&=&{}_{LO}\langle\vec{p}\,'\text{-}|V_{NLO}|\vec{p}+\rangle_{LO}
\end{eqnarray}
with each term represented by a diagram in the second line in Fig.~\ref{fig_bubble}. Substituted in Eq.~\ref{eq_LST_NLO} yields Eq.~\ref{eq_wfkt_NLO}.
\vspace{0.75cm}
\section*{Acknowledgments}
JK gratefully acknowledges the hospitality of the Ohio State University,
discussions with N.~Barnea, H.~Deleon, R. J.~Furnstahl, U.~van Kolck,
S.~K\"onig, J.~Vanasse, and the financial support of the Minerva Foundation.
DG was supported, in part, by BMBF ARCHES.
Both authors are grateful for U.~van Kolck's and B.~Bazak's comments on the manuscript.
%
\bibliographystyle{unsrt}
\bibliography{PILESS_NLO_arX_r1}

\begin{thebibliography}{10}

\bibitem{1979PhyA...96..327W}
S.~{Weinberg}.
\newblock {Phenomenological Lagrangians}.
\newblock {\em Physica A Statistical Mechanics and its Applications},
  96:327--340, April 1979.

\bibitem{1990PhLB..251..288W}
S.~{Weinberg}.
\newblock {Nuclear forces from chiral lagrangians}.
\newblock {\em Physics Letters B}, 251:288--292, November 1990.

\bibitem{1991NuPhB.363....3W}
S.~{Weinberg}.
\newblock {Effective chiral lagrangians for nucleon-pion interactions and
  nuclear forces}.
\newblock {\em Nuclear Physics B}, 363:3--18, September 1991.

\bibitem{1984AnPhy.158..142G}
J.~{Gasser} and H.~{Leutwyler}.
\newblock {Chiral perturbation theory to one loop}.
\newblock {\em Annals of Physics}, 158:142--210, November 1984.

\bibitem{Bethe146}
H.~Bethe and R.~Peierls.
\newblock Quantum theory of the diplon.
\newblock {\em Proceedings of the Royal Society of London A: Mathematical,
  Physical and Engineering Sciences}, 148(863):146--156, 1935.

\bibitem{Bethe176}
H.~A. Bethe and R.~Peierls.
\newblock The scattering of neutrons by protons.
\newblock {\em Proceedings of the Royal Society of London A: Mathematical,
  Physical and Engineering Sciences}, 149(866):176--183, 1935.

\bibitem{PhysRev.76.38}
H.~A. Bethe.
\newblock Theory of the effective range in nuclear scattering.
\newblock {\em Phys. Rev.}, 76:38--50, Jul 1949.

\bibitem{vanKolck1999273}
U.~van Kolck.
\newblock Effective field theory of short-range forces.
\newblock {\em Nuclear Physics A}, 645(2):273 -- 302, 1999.

\bibitem{Kaplan:1998tg}
D.~B. Kaplan, M.~J. Savage, and M.~B. Wise.
\newblock {A New expansion for nucleon-nucleon interactions}.
\newblock {\em Phys. Lett.}, B424:390--396, 1998.

\bibitem{Birse1999169}
M.~C. Birse, J.~A. McGovern, and Keith~G Richardson.
\newblock A renormalisation-group treatment of two-body scattering.
\newblock {\em Physics Letters B}, 464(3–4):169 -- 176, 1999.

\bibitem{PhysRevLett.82.463}
P.~F. Bedaque, H.-W. Hammer, and U.~van Kolck.
\newblock Renormalization of the three-body system with short-range
  interactions.
\newblock {\em Phys. Rev. Lett.}, 82:463--467, Jan 1999.

\bibitem{1970PhLB...33..563E}
V.~{Efimov}.
\newblock {Energy levels arising from resonant two-body forces in a three-body
  system}.
\newblock {\em Physics Letters B}, 33:563--564, December 1970.

\bibitem{Bedaque:1999ve}
P.~F. Bedaque, H.~W. Hammer, and U.~van Kolck.
\newblock {Effective theory of the triton}.
\newblock {\em Nucl. Phys.}, A676:357--370, 2000.

\bibitem{Griesshammer:2005ga}
H.~W. Grie{\ss}hammer.
\newblock {Naive dimensional analysis for three-body forces without pions}.
\newblock {\em Nucl. Phys.}, A760:110--138, 2005.

\bibitem{Kong:1999tw}
X.~Kong and F.~Ravndal.
\newblock {Proton proton fusion in leading order of effective field theory}.
\newblock {\em Nucl. Phys.}, A656:421--429, 1999.

\bibitem{PhysRev.75.1637}
G.~F. Chew and M.~L. Goldberger.
\newblock On the analysis of nucleon-nucleon scattering experiments.
\newblock {\em Phys. Rev.}, 75:1637--1644, Jun 1949.

\bibitem{RevModPhys.22.77}
J.~D. Jackson and J.~M. Blatt.
\newblock The interpretation of low energy proton-proton scattering.
\newblock {\em Rev. Mod. Phys.}, 22:77--118, Jan 1950.

\bibitem{Note1}
This term is small relative to its analog in the neutron-proton channels which
  reconciles the two statements, eventually.

\bibitem{Kong2000137}
X.~Kong and F.~Ravndal.
\newblock Coulomb effects in low energy proton-proton scattering.
\newblock {\em Nuclear Physics A}, 665(1-2):137 -- 163, 2000.

\bibitem{PhysRevLett.29.917}
M.~E. Fisher, S.-k. Ma, and B.~G. Nickel.
\newblock Critical exponents for long-range interactions.
\newblock {\em Phys. Rev. Lett.}, 29:917--920, Oct 1972.

\bibitem{PhysRevC.67.064006}
T.~Barford and M.~C. Birse.
\newblock Renormalization group approach to two-body scattering in the presence
  of long-range forces.
\newblock {\em Phys. Rev. C}, 67:064006, Jun 2003.

\bibitem{2010JPhG...37j5108A}
S.-i. {Ando} and M.~C. {Birse}.
\newblock {Effective field theory of $^{3}$He}.
\newblock {\em Journal of Physics G Nuclear Physics}, 37(10):105108, October
  2010.

\bibitem{PhysRevC.83.064001}
S.~K{\"o}nig and H.-W. Hammer.
\newblock Low-energy $p$-$d$ scattering and $^{3}\mathrm{He}$ in pionless
  effective field theory.
\newblock {\em Phys. Rev. C}, 83:064001, Jun 2011.

\bibitem{FewBodySyst.54.1}
S.~K\"onig and H.-W. Hammer.
\newblock The low-energy p-d system in pionless eft.
\newblock {\em Few-Body Systems}, 54(1-4):231--234, 2013.

\bibitem{PhysRevC.89.064003}
J.~Vanasse, D.~A. Egolf, J.~Kerin, S.~K{\"o}nig, and R.~P. Springer.
\newblock $^{3}\mathrm{He}$ and $pd$ scattering to next-to-leading order in
  pionless effective field theory.
\newblock {\em Phys. Rev. C}, 89:064003, Jun 2014.

\bibitem{vanKolck:1998bw}
U.~van Kolck.
\newblock {Effective field theory of short range forces}.
\newblock {\em Nucl. Phys.}, A645:273--302, 1999.

\bibitem{Note2}
See~\cite {Rotureau:2011vf} for a visualization of the different scalings.

\bibitem{Hammer:2007kq}
H.~W. Hammer, D.~R. Phillips, and L.~Platter.
\newblock {Pion-mass dependence of three-nucleon observables}.
\newblock {\em Eur. Phys. J.}, A32:335--347, 2007.

\bibitem{Stetcu:2006ey}
I.~Stetcu, B.~R. Barrett, and U.~van Kolck.
\newblock {No-core shell model in an effective-field-theory framework}.
\newblock {\em Phys. Lett.}, B653:358--362, 2007.

\bibitem{Wiringa:2013ala}
R.~B. Wiringa, R.~Schiavilla, Steven~C. Pieper, and J.~Carlson.
\newblock {Nucleon and nucleon-pair momentum distributions in $A \le 12$
  nuclei}.
\newblock {\em Phys. Rev.}, C89(2):024305, 2014.

\bibitem{PhysRevLett.49.974}
E.~Jans, P.~Barreau, M.~Bernheim, J.~M. Finn, J.~Morgenstern, J.~Mougey,
  D.~Tarnowski, S.~Turck-Chieze, S.~Frullani, F.~Garibaldi, G.~P. Capitani,
  E.~de~Sanctis, M.~K. Brussel, and I.~Sick.
\newblock Quasifree ($e, {e}^{\ensuremath{'}}p$) reaction on $^{3}\mathrm{He}$.
\newblock {\em Phys. Rev. Lett.}, 49:974--978, Oct 1982.

\bibitem{Note3}
\cite {Wiringa:2013ala,PhysRevLett.49.974} extract occupation numbers. However,
  it was shown~\cite {Furnstahl:2001xq} that occupation numbers cannot be
  defined uniquely. We therefore interpret the large contribution of
  low-momentum modes of~\cite {Wiringa:2013ala,PhysRevLett.49.974} as specific
  to the employed interaction and extraction method.

\bibitem{Bedaque:2002yg}
P.~F. Bedaque, G.~Rupak, H.~W. Grie{\ss}hammer, and H.-W. Hammer.
\newblock {Low-energy expansion in the three-body system to all orders and the
  triton channel}.
\newblock {\em Nucl. Phys.}, A714:589--610, 2003.

\bibitem{Note4}
See Bethe's original ERE expansion about the deuteron pole~\cite
  {PhysRev.76.38} or the fixing of the residue in~\cite {Phillips:1999hh} as
  examples of wise choices for the momenta to expand about.

\bibitem{PhysRevC.26.2381}
L.~P. Kok, J.~W. de~Maag, H.~H. Brouwer, and H.~van Haeringen.
\newblock Formulas for the $\ensuremath{\delta}$-shell-plus-coulomb potential
  for all partial waves.
\newblock {\em Phys. Rev. C}, 26:2381--2396, Dec 1982.

\bibitem{Note5}
As one presumably given in~\cite {Konig:2015aka}, where the authors address the
  same complex as this article does, and of which we became aware while
  finalizing this manuscript.

\bibitem{Note6}
Here, this value is apparently $<400~$MeV.

\bibitem{nocedal1999numerical}
J.~Nocedal and S.J. Wright.
\newblock {\em Numerical Optimization}.
\newblock Springer Series in Operations Research. Springer, 1999.

\bibitem{0953-4075-36-19-013}
H.~Ouerdane, M.~J. Jamieson, D.~Vrinceanu, and M.~J. Cavagnero.
\newblock The variable phase method used to calculate and correct scattering
  lengths.
\newblock {\em Journal of Physics B: Atomic, Molecular and Optical Physics},
  36(19):4055, 2003.

\bibitem{Wheeler:1937zza}
J.~A. Wheeler.
\newblock {Molecular Viewpoints in Nuclear Structure}.
\newblock {\em Phys. Rev.}, 52:1083--1106, 1937.

\bibitem{Wheeler:1937zz}
J.~A. Wheeler.
\newblock {On the Mathematical Description of Light Nuclei by the Method of
  Resonating Group Structure}.
\newblock {\em Phys. Rev.}, 52:1107--1122, 1937.

\bibitem{hmh-rrgm}
H.M. Hofmann.
\newblock In L.S. Ferreira, A.C. Fonseca, and L.~Streit, editors, {\em
  Proceedings of Models and Methods in Few-Body Physics, Lisboa, Portugal},
  page 243, 1986.

\bibitem{Kirscher:2011zn}
J.~Kirscher and D.~R. Phillips.
\newblock {Constraining the neutron-neutron scattering length using the
  effective field theory without explicit pions}.
\newblock {\em Phys. Rev.}, C84:054004, 2011.

\bibitem{Note7}
As the $\protect \mathbb {R}$GM~is readily replaced with another method, all
  necessary techniques are standard and the described implementation can easily
  be reproduced.

\bibitem{Furnstahl:2014xsa}
R.~J. Furnstahl, D.~R. Phillips, and S.~Wesolowski.
\newblock {A recipe for EFT uncertainty quantification in nuclear physics}.
\newblock {\em J. Phys.}, G42(3):034028, 2015.

\bibitem{Epelbaum:2014sza}
E.~Epelbaum, H.~Krebs, and U.~G. Mei{\ss}ner.
\newblock {Precision nucleon-nucleon potential at fifth order in the chiral
  expansion}.
\newblock {\em Phys. Rev. Lett.}, 115(12):122301, 2015.

\bibitem{Epelbaum:2002gb}
E.~Epelbaum, U.-G. Mei{\ss}ner, and W.~Gl{\"o}ckle.
\newblock {Nuclear forces in the chiral limit}.
\newblock {\em Nucl. Phys.}, A714:535--574, 2003.

\bibitem{Kirscher:2009aj}
J.~Kirscher, H.~W. Grie{\ss}hammer, D.~Shukla, and H.~M. Hofmann.
\newblock {Universal Correlations in Pion-less EFT with the Resonating Group
  Model: Three and Four Nucleons}.
\newblock {\em Eur. Phys. J.}, A44:239--256, 2010.

\bibitem{Platter:2004zs}
L.~Platter, H.~W. Hammer, and Ulf-G. Mei{\ss}ner.
\newblock {On the correlation between the binding energies of the triton and
  the alpha-particle}.
\newblock {\em Phys. Lett.}, B607:254--258, 2005.

\bibitem{Nogga:2005hy}
A.~Nogga, R.~G.~E. Timmermans, and U.~van Kolck.
\newblock {Renormalization of one-pion exchange and power counting}.
\newblock {\em Phys. Rev.}, C72:054006, 2005.

\bibitem{Long:2011xw}
B.~Long and C.~J. Yang.
\newblock {Renormalizing Chiral Nuclear Forces: Triplet Channels}.
\newblock {\em Phys. Rev.}, C85:034002, 2012.

\bibitem{Long:2012ve}
B.~Long and C.~J. Yang.
\newblock {Short-range nuclear forces in singlet channels}.
\newblock {\em Phys. Rev.}, C86:024001, 2012.

\bibitem{Valderrama:2011mv}
M.~P. Valderrama.
\newblock {Perturbative Renormalizability of Chiral Two Pion Exchange in
  Nucleon-Nucleon Scattering: P- and D-waves}.
\newblock {\em Phys. Rev.}, C84:064002, 2011.

\bibitem{Valderrama:2009ei}
M.~P. Valderrama.
\newblock {Perturbative renormalizability of chiral two pion exchange in
  nucleon-nucleon scattering}.
\newblock {\em Phys. Rev.}, C83:024003, 2011.

\bibitem{Note8}
\cite {Baru:2015ira} is a recent example for reinterpreting a methodology
  originally devised for long-range Coulomb and short-range nuclear
  interactions through an identification of the pion-exchange as long ranged.

\bibitem{Taylor1972}
J.~R. Taylor.
\newblock {\em {Scattering Theory}}.
\newblock Wiley, 1972.

\bibitem{Rotureau:2011vf}
J.~Rotureau, I.~Stetcu, B.~R. Barrett, and U.~van Kolck.
\newblock {Two and Three Nucleons in a Trap and the Continuum Limit}.
\newblock {\em Phys. Rev.}, C85:034003, 2012.

\bibitem{Furnstahl:2001xq}
R.~J. Furnstahl and H.~W. Hammer.
\newblock {Are occupation numbers observable?}
\newblock {\em Phys. Lett.}, B531:203--208, 2002.

\bibitem{Phillips:1999hh}
D.~R. Phillips, G.~Rupak, and M.~J. Savage.
\newblock {Improving the convergence of N N effective field theory}.
\newblock {\em Phys. Lett.}, B473:209--218, 2000.

\bibitem{Konig:2015aka}
S.~K{\"o}nig, H.~W. Grie{\ss}hammer, H.~W. Hammer, and U.~van Kolck.
\newblock {Effective Theory of 3H and 3He}.
\newblock 2015.

\bibitem{Baru:2015ira}
V.~Baru, E.~Epelbaum, A.~A. Filin, and J.~Gegelia.
\newblock {Low-energy theorems for nucleon-nucleon scattering at unphysical
  pion masses}.
\newblock {\em Phys. Rev.}, C92(1):014001, 2015.

\end{thebibliography}
\end{document}